\documentclass[showpacs,pra,twocolumn]{revtex4}

\usepackage{epsfig,amsmath}
\usepackage{subfigure}
\usepackage{graphicx}
\usepackage{dcolumn}
\usepackage{stmaryrd}
\usepackage{mathrsfs}
\usepackage{pifont}
\usepackage{amsthm}
\usepackage{amssymb}
\usepackage{bm}
\usepackage{latexsym}
\usepackage{hyperref}
\usepackage{color}
\usepackage{appendix}
\usepackage{braket}
\usepackage{amsmath}
\usepackage{diagbox}
\usepackage{algorithm} 
\usepackage{algpseudocode} 

\newcommand{\beq}{\begin{equation}}
	\newcommand{\eeq}{\end{equation}}
\newcommand{\beqa}{\begin{eqnarray}}
	\newcommand{\eeqa}{\end{eqnarray}}

\begin{document}
	
	\title{Arbitrary quantum states preparation aided by deep reinforcement learning}
	
	\begin{abstract}
	The preparation of quantum states is essential in the realm of quantum information processing, and the development of efficient methodologies can significantly alleviate the strain on quantum resources. Within the framework of deep reinforcement learning (DRL), we integrate the initial and the target state information within the state preparation task together, so as to realize the control trajectory design between two arbitrary quantum states.  Utilizing a semiconductor double quantum dots (DQDs) model, our results demonstrate that the resulting control trajectories can effectively achieve arbitrary quantum state preparation (AQSP) for both single-qubit and two-qubit systems, with average fidelities of 0.9868 and 0.9556 for the test sets, respectively. Furthermore, we consider the noise around the system and the control trajectories exhibit commendable robustness against charge and nuclear noise. Our study not only substantiates the efficacy of DRL in QSP, but also provides a new solution for quantum control tasks of multi-initial and multi-objective states, and is expected to be extended to a wider range of quantum control problems.
	\end{abstract}

	\author{Zhao-Wei Wang$^{1}$ and Zhao-Ming Wang$^{1}$\footnote{wangzhaoming@ouc.edu.cn}}
	\affiliation{$^{1}$ College of Physics and Optoelectronic Engineering, Ocean University of China, Qingdao 266100, China}
	\maketitle

	\section{Introduction.}

    Precise control of quantum dynamics in physical systems is a cornerstone of quantum information processing. Achieving high-fidelity quantum state preparation (QSP) is crucial for quantum computing and simulation \cite{nielsen2010quantum,cho2021quantum}. This process often requires iterative solutions of a set of nonlinear equations \cite{krauss2023optimizing,wang2014robust, wang2012composite,throckmorton2017fast}, which is complex and time-consuming. Consequently, the quest for efficient methods to prepare arbitrary quantum states has become a prominent issue in quantum control.

	In this context, protocols based on quantum optimal control theory \cite{ferrie2014self, doria2011optimal,khaneja2005optimal} have been acquiring increasing attention. Traditional gradient-based optimization methods, such as stochastic gradient descent (SGD) \cite{ferrie2014self}, chopped random-basis optimization (CRAB) \cite{doria2011optimal,caneva2011chopped}, and gradient ascent pulse engineering (GRAPE) \cite{khaneja2005optimal,rowland2012implementing}, have been employed to address optimization challenges. However, these methods tend to produce nearly continuous pulses, which may not be ideal for experimental implementation. Recently, machine learning techniques, such as deep reinforcement learning (DRL) has emerged as a more efficient approach for designing discrete control pulses, offering lower control costs and significant control effects compared to traditional optimization algorithms \cite{zhang2019does,he2021deep}.

	DRL enhances reinforcement learning with neural networks, significantly boosting the ability of agents to recognize and learn from complex features \cite{mnih2015human}. This enhancement has paved the way for a broad spectrum of applications in quantum physics \cite{yu2022deep,yu2023event,wauters2020reinforcement,dong2008quantum,neema2024non,bukov2018reinforcement,martin2021reinforcement}, including QSP \cite{zhang2019does,he2021deep,chen2013fidelity}, quantum circuit gate optimization \cite{shindi2023model,niu2019universal,an2019deep,baum2021experimental}, coherent state transmission \cite{porotti2019coherent}, adiabatic quantum control \cite{ding2021breaking}, and the measurement of quantum devices \cite{nguyen2021deep}.

	DRL assisted QSP has been studied a lot, such as fixed QSP from specific quantum states to other designated states \cite{liu2022quantum, chen2013fidelity}. Multi-objective control from fixed states to arbitrary states \cite{zhang2019does, haug2020classifying}, or from arbitrary states to fixed state \cite{he2021deep}, also have promising results. Then the arbitrary quantum state preparation (AQSP) can be obtained by combining these two opposite directions of QSP immediately, which involves preparing multi-initial states and multi-objective states \cite{he2021universal}. However, two steps: from arbitrary to fixed state and then to arbitrary states, must be required in above stategy. Can we use a unified method to realize AQSP from abitray states to arbitray states directly? In this paper, we successfully accomplish the task of designing control trajectories for the AQSP by harnessing the DRL algorithm to condense the initial and target quantum state information into a unified representation. We also incorporate the positive-operator valued measure (POVM) method \cite{carrasquilla2019reconstructing,carrasquilla2021probabilistic,luo2022autoregressive, reh2021time} to address the complexity of the density matrix elements that are not readily applicable in machine learning. Take the semiconductor double quantum dots (DQDs) model as a testbed, we assess the performance of our algorithm-designed action trajectories in the AQSP for both single-qubit and two-qubit. At last, we consider the effectiveness of our algorithm on the noise problme and the results show the robustness of the designed control trajectories against charge noise and nuclear noise.

	\section{MODEL}
	
	In the architecture of a circuit-model quantum computer, the construction of any quantum logic gate is facilitated through the combination of single-qubit gates and entangled two-qubit gates \cite{nielsen2010quantum}.  This study focuses on the AQSP for both single-qubit and two-qubit.  We have adopted the spin singlet-triplet $(S-T_0)$ encoding scheme within DQDs for qubit encoding, a method that is favored for its ability to be manipulated solely by electrical pulses \cite{taylor2005fault,nichol2017high,wu2014two}.

	The spin singlet state and the spin triplet state are encoded as $|0\rangle=|S\rangle=(|\uparrow\downarrow\rangle-|\downarrow\uparrow\rangle)/\sqrt{2}$ and
	$|1\rangle=|T_0\rangle=(|\uparrow\downarrow\rangle+|\downarrow\uparrow\rangle)/\sqrt{2}$, respectively. 
	Here $|\uparrow\rangle$ and $|\downarrow\rangle$ denote the two spin eigenstates of a single electron. 
	The control Hamiltonian for a single-qubit in the semiconductor DQDs model is given by \cite{malinowski2017notch,maune2012coherent}:
	\begin{equation}
		H(t)=J(t)\sigma_z+h\sigma_x,
		\label{equ:1}
	\end{equation}
	where $\sigma_z$ and $\sigma_x$ are the Pauli matrix components in the $z$ and $x$ directions, respectively.
	$J(t)$ is a positive, adjustable parameter, while $h$ symbolizes the Zeeman energy level separation between two spins, typically regarded as a constant \cite{zhang2019semiconductor}. For simplicity, we set $h=1$ and use the reduced Planck constant $\hbar=1$ throughout.

	In the field of quantum information processing, operations on entangled qubits are indispensable. Within semiconductor DQDs, interqubit operations are performed on two adjacent $S-T_0$ qubits that are capacitively coupled. The Hamiltonian, in the basis of $\{|SS\rangle,|T_0S\rangle,|ST_0\rangle,|T_0T_0\rangle\}$, is expressed as \cite{taylor2005fault,nichol2017high,shulman2012demonstration,van2011charge}:
	\begin{equation}
	\begin{aligned}
		H(t)=
		&\frac{1}{2}\Big(J_{1}\left(\sigma_{z}\otimes I\right)\:+\:J_{2}(I\otimes\sigma_{z})\\
		&+\frac{J_{12}}{2}\left(\left(\sigma_{z}-I\right)\otimes\left(\sigma_{z}-I\right)\right)\\
		&+h_1(\sigma_{x}\otimes I)+h_2(I\otimes\sigma_{x})\Big),
	\end{aligned}
	\label{equ:2}
	\end{equation}
	where $J_i$ and $h_i$ represent the exchange coupling and the Zeeman energy gap of the $i$th qubit, respectively. $J_{12}$ is proportional to $J_1J_2$, representing the magnitude of the coulomb coupling between the two qubits. It is crucial for $J_i$ to be positive to maintain consistent interqubit coupling. To streamline the model, we assume $h_1 = h_2 = 1$ and set $J_{12} = J_1J_2/2$ in this context.

	\section{METHODS}
	\subsection{Positive-Operator Valued Measure (POVM)}
	
	Typically, the density matrix elements $\rho_{ij}$ are complex numbers. However, standard machine learning algorithms can not be used to handle complex numbers directly. To solve this problem, a straightforward approach is to decompose each complex number into its real and imaginary components, reorganize them into a new data following specific protocols, and subsequently input this data into the machine learning model. Once processed, the data can be reassembled into their original complex form using the inverse of the initial transformation rules. 	Beyond this method, recent advancements in applying machine learning to quantum information tasks have used POVM method to deal with the  complex number problems of the density matrix \cite{carrasquilla2019reconstructing,carrasquilla2021probabilistic,luo2022autoregressive, reh2021time}. Specifically, a collection of positive semi-definite measurement operators $\mathbf{M}=\{M_{(\mathbf{a})}\}$ is utilized to translate the density matrix into a corresponding set of measurement outcomes $\mathbf{P}=\{P_{(\mathbf{a})}\}$. When these outcomes fully capture the information content of the density matrix, they are termed informationally complete POVMs (IC-POVMs). These operators adhere to the normalization condition $\sum_\mathbf{a}M_{(\mathbf{a})}=\mathbb{I}$.

	For an N-qubit system, the density matrix can be converted to the form of a probability distribution by 
	\begin{equation}
		P_{(\mathbf{a})}=\mathrm{tr}[\rho M_{(\mathbf{a})}],
		\label{equ:3}
	\end{equation}
	where $M_{(\mathbf{a})} = M_{(a_1)}\otimes..\otimes M_{(a_N)}$. 
	In this work, we employ the Pauli-4 POVM 
	$\mathbf{M}_{\mathrm{Pauli-4}}= \{ 
	M_{(a_i)}^1=\frac{1}{3}\times|0\rangle\langle 0|,\ 
	M_{(a_i)}^2=\frac{1}{3}\times|l\rangle\langle l|,\ 
	M_{(a_i)}^3=\frac{1}{3}\times|+\rangle\langle +|,\ 
	M_{(a_i)}^4=\mathbb{I}-	M_{(a_i)}^1-M_{(a_i)}^2-M_{(a_i)}^3\}.$
	By inverting Eq.~(\ref{equ:3}), we can retrieve the density matrix $\rho$ as follows:
	\begin{equation}
		\rho=\sum_{\mathbf{a}}\sum_{\mathbf{a}^{\prime}}P(\mathbf{a})T^{-1}_{\mathbf{aa}^{\prime}}M({\mathbf{a}^{\prime}}),
		\label{equ:4}
	\end{equation}
	where $T_{\mathbf{aa}^{\prime}} = \operatorname{tr}(M_{(\mathbf{a})}M_{({\mathbf{a}^{\prime}})})$ represents an element of the overlap matrix $T$. More details of the POVM, see Ref.~ \cite{carrasquilla2019reconstructing}.

	\subsection{Arbitrary quantum states preparation via DRL}

	Our objective is to accomplish AQSP using discrete rectangular pulses \cite{khaneja2005optimal,wang2012composite,wang2014robust,krantz2019quantum, shulman2012demonstration}. To this end, we use the Deep Q-Network (DQN) algorithm \cite{mnih2013playing, mnih2015human}, which is one of the important methods of DRL, to formulate action trajectories. Details of DQN algorithm are put in Appendix A.
	
	At first, we sample uniformly on the surface of the Bloch sphere to identify the initial quantum states $\rho_{ini}$ and the target quantum states $\rho_{tar}$ for the QSP. Then we construct a data set for training, validation, and test.
	In the context of universal state preparation (USP) \cite{he2021deep}, which involves transitioning from any arbitrary $\rho_{ini}$ to a predetermined $\rho_{tar}$, the data set is compiled solely with $\rho_{ini}$ instances, using the fixed $\rho_{tar}$ to assess the efficacy of potential actions. For tasks involving the preparation of diverse $\rho_{tar}$ states, supplementary network training is required.

	In scenarios that demand handling multiple initial states and objectives, such as AQSP, our aim is to enable the Agent to discern among various $\rho_{tar}$ states and to devise the corresponding control trajectories. Thus, in the process of data set design, we take the information from $\rho_{ini}$ and $\rho_{tar}$ to form the state $s$ within the DQN algorithm, expressed as $s=[P_{ini}^1,...,P_{ini}^n, P_{tar}^1,...,P_{tar}^n]$. Here, the POVM method is employed to transform the density matrix $\rho$ into a probability distribution $\{P_{(\mathbf{a})}\}$. The first segment of $s$ primarily serves the evolutionary computations, while the latter portion is utilized to distinguish between different tasks and to compute the reward values associated with actions. The data set is then randomly shuffled and partitioned into training, validation, and test subsets. The training set is predominantly used for the Main Net's training, the validation set assists in estimating the generalization error during the training phase, and the test set is employed to assess the Main Net's performance post-training.

	Subsequently, the Main Net, initialized at random, samples the input state $s$ from the training set at each step $k = 1$ and subsequently predicts the optimal action $a_k$ (i.e., the pulse intensity $J(t)$). From $s$, the sets $\{P_{k}\}$ and $\{P_{tar}\}$ are isolated, and the states $\rho_{k}$ and $\rho_{tar}$ are calculated using Eq.~(\ref{equ:4}). Given the current $\rho_{k}$ and the chosen action $a_k$, we compute the next state $\rho_{k}'=- i[H{(a_k)},\rho_{k}]$ and determine the fidelity $F(\rho_{tar},\rho_{k})\:\equiv\:\mathrm{Tr}\left(\sqrt{\sqrt{\rho_{tar}}\rho_{k}\sqrt{\rho_{tar}}}\right)$. The fidelity $F$ serves as a critical metric, quantifying the proximity between the subsequent state and the target state. Utilizing Eq.~(\ref{equ:3}), we derive the set $\{P_{k}'\}$, which, when merged with $\{P_{tar}\}$ , allows us to reconstruct the new state $s'$. This updated state $s'$ is then introduced to the Main Network as the current state $s$, with the iteration index $k$ incremented by one. The reward value $r=r (F)$, which is instrumental in training the Main Network, is formulated as a function of fidelity. This process is iteratively executed until the iteration count $k$ reaches its maximum limit or the task completion criteria are satisfied, indicated by $F>F_{threshold}$. 
	After completing the action sequence designed by AQSP algorithm, the initial state $\rho_{ini}$ evolves to the final state $\rho_{fin}$. We use the fidelity $F$ of the final state $\rho_{fin}$ and the target state $\rho_{tar}$ to judge the quality of the action sequence. The average fidelity is the average of the fidelity of all tasks in the entire data set.
	Ultimately, following extensive training, the Main Net attains the capability to assign a $Q$-value to each state-action pair. With precise $Q$-values at our disposal, we can determine an appropriate action for any given state, including those that were not explicitly trained.

	A complete description of the training process and the data format conversion is delineated in Algorithm \ref{alg:1}, which we define it as AQSP algorithm. For the computational aspects of the algorithm, we employed the Quantum Toolbox in Python \cite{johansson2012qutip}.

	\begin{algorithm}
		\caption{The pseudocode for training the AQSP algorithm}
		\begin{algorithmic}[1]
			\State \textbf{Initialize} the Experience Memory $D$.
			\State \textbf{Initialize} the Main-network $\theta$.
			\State \textbf{Initialize} the Target-network $\theta^-$ by: $\theta^- \gets \theta$.
			\State \textbf{Set} the $\epsilon = 0$.
			\For{episode=0, $\mathrm{episode}_{\mathrm{max}}$}
			
			\For{Select state $s = s_{input}$ from the training set}
			\State Split $P$ and $P_{tar}$ from $s$.
			\State Convert $P(P_{tar})$ to $\rho (\rho_{tar})$ by the inverse operation of POVM.
			
			\While{True}
			\State With probability $1-\epsilon $ select a random action, otherwise $a_i=\mathrm{argmax}_aQ(s,a;\theta)$.
			\State Set the $\epsilon=\epsilon+\delta\epsilon$, except$\epsilon=\epsilon_{max}$.
			\State Perform $a_i$ and get the next quantum state $\rho^{\prime}$.
			\State Calculate fidelity $F$ and obtain the reward $r$.
			\State Convert $\rho^{\prime}$ to $P^{\prime}$ with POVM.
			\State Concatenate $P^{\prime}$ and $P_{tar}$ to $s^{\prime}$.
			\State Store experience unit $(s, a_i, r , s^{\prime})$ in $D$.
			\State Select Batch size $N_{bs}$ of experiences units randomly from $D$ for training.
			\State Update $\theta$ by minimizing the $Loss$ function.
			\State Every $C$ steps,  $\theta^- \gets \theta$.
			\State $\mathbf{break}\mathrm{~if~}F>F_{\mathrm{threshold}}\mathrm{~or~step}\geq T/\mathrm{d}t.$
			\EndWhile
			
			\EndFor
			
			\EndFor
			
		\end{algorithmic}
		\label{alg:1}
	\end{algorithm}

	\section{RESULTS AND DISCUSSIONS}
	\subsection{Single-qubit}

	\begin{table}[htbp]
		\centering
		\caption{Hyperparameter table}
		
	\begin{tabular}{lll}
		\hline
		Qubit quantity & Single-qubit & Two-qubit \\
		\hline
		Allowed action $a(J(t))$ & $0,1,2,3,4$ & $\{(i,j)\}^*$ \\
		Size of the training set & $100$ & $100$    \\
		Size of the validation set & $100$ & $100$   \\
		Size of the test set & $9506$ & $39600$    \\
		Batch size $N_{bs}$ & $32$ & $32$    \\
		Memory size $M$ & $20000$ & $30000$    \\
		Learning rate $\alpha $ & $0.001$ & $0.001$    \\
		Replace period $C$ & $S200$ & $200$    \\
		Reward discount factor $\gamma$ & $0.9$ & $0.9$    \\
		Number of hidden layers & $2$ & $3$    \\
		Neurons per hidden layer & $64/64$ & $128/128/64$    \\
		Activation function & Relu & Relu    \\
		$\epsilon$-greedy increment $\delta\epsilon$  & $0.001$ & $0.0001$    \\
		Maximal $\epsilon$ in training $\epsilon_{max}$  & $0.95$ & $0.95$    \\
		$\epsilon$ in validation and test  & $1$ & $1$    \\
		$F_{threshold}$ per episode & $0.999$ & $0.99$    \\
		Episode for training  & $100$ & $400$    \\
		Total time $T$  & $2\pi$ & $5\pi$    \\
		Action duration $\mathrm{d}t$  & $\pi/5$ & $\pi/4$    \\
		Maximum steps per episode  & $10$ & $20$    \\
		\hline
		$\{(i,j)\ \ i,j=0,1,2,3,4,5\}^* $
		
	\end{tabular}
		\label{tab:1}
		
	\end{table}

	We now employ our AQSP algorithm to achieve the task of preparing a single-qubit from an arbitrary initial state to an arbitrary target state. To this end, we sample $98$ quantum states uniformly across the Bloch sphere, varying the parameters $\alpha$ and $\beta$, and subsequently construct a data set comprising $98*97=9506$ data points by pairing each state as both the initial and target states. The training and validation sets each contain $100$ data points, with the remaining $9306$ data points allocated to the test set. We define five distinct actions $J (t)$ with values $0, 1, 2, 3$ and $4$. Each action pulse has a duration $dt=\pi/5$, the total evolution time is $2\pi$, and a maximum of $10$ actions are allowed per task. The Main Network is structured with two hidden layers, each consisting of $64$ neurons. The reward function is set as $r=F$, with all algorithm hyperparameters detailed in Table~\ref{tab:1}.

\begin{figure}
	\centerline{\includegraphics[width=1.0\columnwidth]{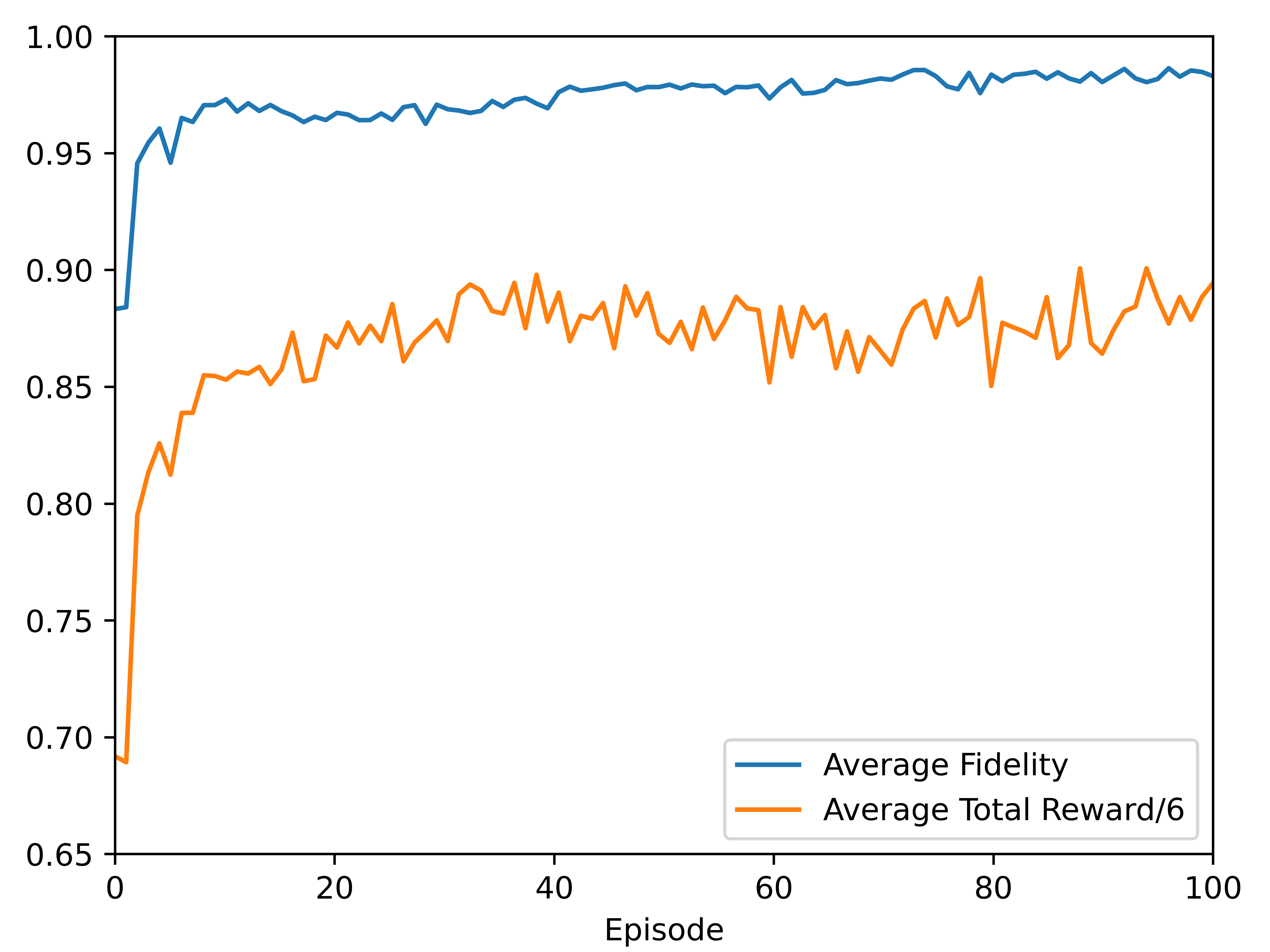}}
	\caption{The average fidelity and total reward over the validation set as functions of the number of episodes in the training process for the single-qubit AQSP. The average fidelity of the test set is $\bar{F}=0.9864$.}
	\label{fig:1}
	\end{figure}

	As shown in Fig~\ref{fig:1}, the average fidelity and total reward surge rapidly during the initial $6$ training episodes, experience a slight fluctuation, and then plateau around the $77$th episode, signaling the convergence of the Main Network for the AQSP task.

	To illustrate the efficacy of our algorithm, we contrast it with the USP algorithm \cite{he2021deep}. Specifically, we train a model using the USP algorithm with the same hyperparameters to design action trajectories from arbitrary quantum states to the state $\left|1\right\rangle $. For comparative analysis, we replace the original AQSP test set with one where the target state $\rho$ is $\left|1\right\rangle $ during testing. We then assessed the performance of both algorithms in the task of preparing any quantum state to the state $\left|1\right\rangle $. Table \ref{tab:2} records the fidelity (average fidelity) of these different tests. Although the USP algorithm achieves a high average fidelity $\bar{F}=0.9972$ for tasks with the target state $\left|1\right\rangle $, it fails when the target state is $\left|0\right\rangle $. In contrast, the AQSP algorithm demonstrates adaptability across different tasks, yielding favorable outcomes.

\begin{figure}
		\centerline{\includegraphics[width=1.0\columnwidth]{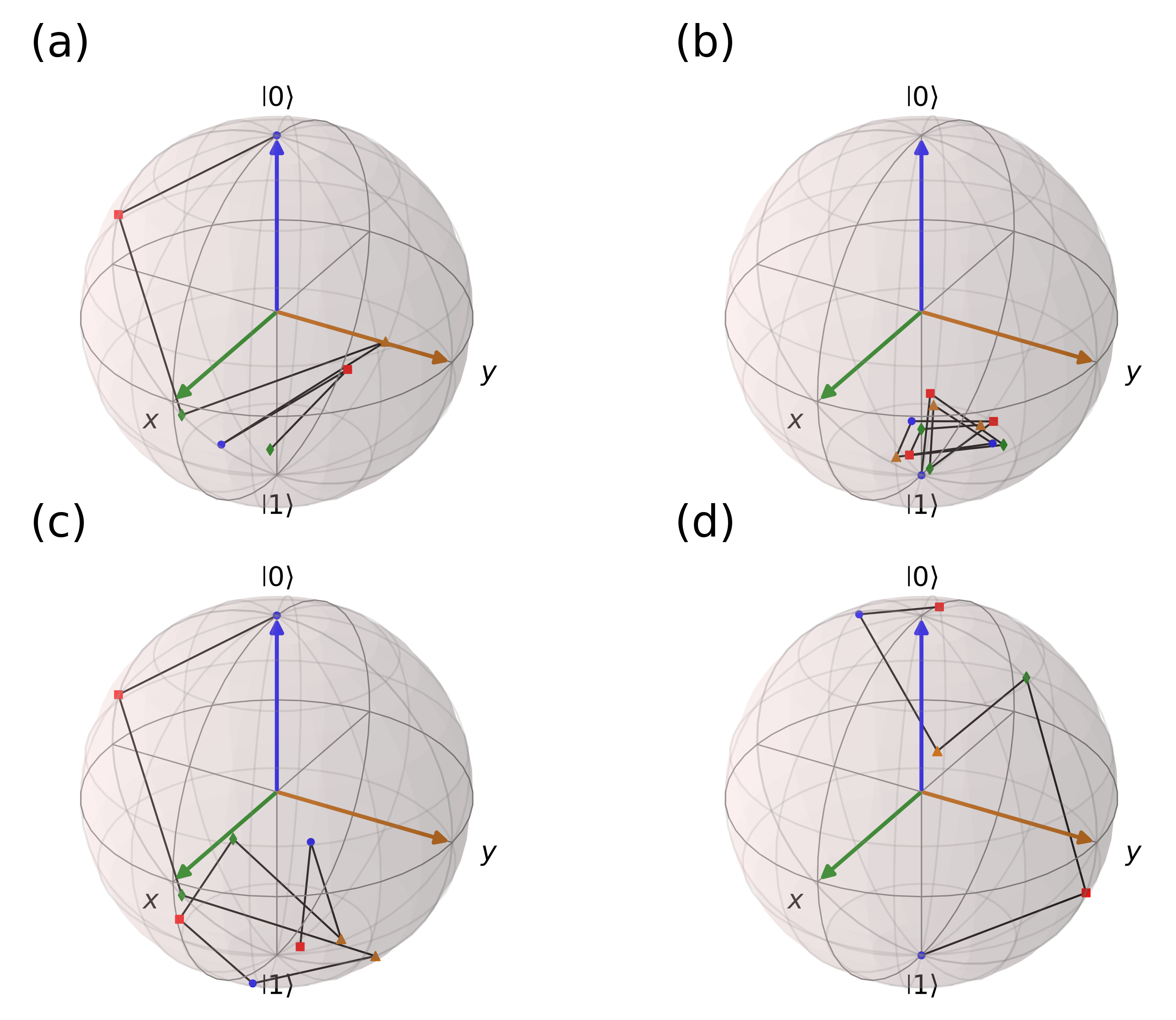}}
		\caption{Motion trajectories designed by AQSP algorithm and USP algorithm in two preparation tasks. (a) and (b) are the action trajectories designed by USP algorithm in the task $|0\rangle\rightarrow|1\rangle$ and $|1\rangle\rightarrow|0\rangle$, respectively. (c) and (d) are the action trajectories designed by AQSP algorithm in the task $|0\rangle\rightarrow|1\rangle$ and $|1\rangle\rightarrow|0\rangle$, respectively.}
		\label{fig:2}
	\end{figure}

	To provide a more intuitive comparison of the control action sequences crafted by the two algorithms, we plotted their specific control trajectories for two QSP tasks: $|0\rangle\rightarrow|1\rangle$ and $|1\rangle\rightarrow|0\rangle$ as shown in Fig~\ref{fig:2}. The AQSP algorithm proves to be highly effective in both tasks, whereas the USP algorithm is only capable of completing the first task. This is attributed to the AQSP algorithm's incorporation of the target state within the training process, enabling it to adapt to a variety of QSP tasks. Conversely, the USP algorithm sets the target state as fixed during training, which results in the designed action trajectory being confined to the  selected target state during training, even if the target state changes. Should the USP algorithm be utilized for the second task, an additional model with the target state $\left|0\right\rangle $ must be trained, which would still be incapable of completing the first task.

A

	\begin{table}[htbp]
	\centering
	\caption{Fidelity (average fidelity) table of USP and AQSP in different QSP tasks}
	\begin{tabular}{lllll}
		\hline
		Task
		& $|a\rangle\rightarrow|1\rangle \ \ $ 
		& $|0\rangle\rightarrow|1\rangle \ \ $ 
		& $|1\rangle\rightarrow|0\rangle \ \ $  
		& $|a\rangle\rightarrow|a\rangle \ \ $ \\
		\hline
		USP & 0.9972  & 0.9941 & 0.2892 &  \\
		AQSP & 0.9932 & 0.9975 & 0.9972 & 0.9864 \\
		\hline
	\end{tabular}
	\label{tab:2}
	\end{table}

	\subsection{Two-qubit}

	We now turn our attention to the AQSP for two-qubit. Our data set encompasses $6912$ points, defined as $\{[a_1,a_2,a_3,a_4]^T\}$, where $a_j = bc_j$ and $b\in\{1,i,-1,-i\}$ belongs to the set $\{1,i,-1,-i\}$, representing the phase. Collectively, a set of $c_j$ values defines a point on a four-dimensional unit hypersphere, as described by Eq.~(\ref{equ:5}):
	\begin{equation}
	\begin{cases}c_1=\cos\theta_1,\\[1ex]c_2=\sin\theta_1\cos\theta_2,\\[1ex]c_3=\sin\theta_1\sin\theta_2\cos\theta_3,\\[1ex]c_4=\sin\theta_1\sin\theta_2\sin\theta_3,\end{cases}
	\label{equ:5}
	\end{equation}
	where $\theta_i\in\{\pi/8,\pi/4,3\pi/8\}$. The quantum states corresponding to these points adhere to the normalization condition. For training and test, we randomly selected $200$ points from this database. Subsequently, we randomly chose $200$ quantum states from the database to serve as both the initial and target states for the QSP task, constructing a data set of $200*199=39800$ entries. From this data set, we randomly designated $100$ quantum states for both the training set and the test set, with the remaining $39600$ states allocated for the test set. Our standard pulse strengths for each qubit are given by the set $\{(J_1,J_2)|J_1,J_2\in\{1,2,3,4,5\}\}$. The pulse duration for each action is $dt=\pi/4$, the total evolution time is $5\pi$, and a maximum of $20$ actions are permitted per task. The reward function is also defined as $r=F$. All hyperparameters are detailed in Table~\ref{tab:1}.

\begin{figure}
		\centerline{\includegraphics[width=1.0\columnwidth]{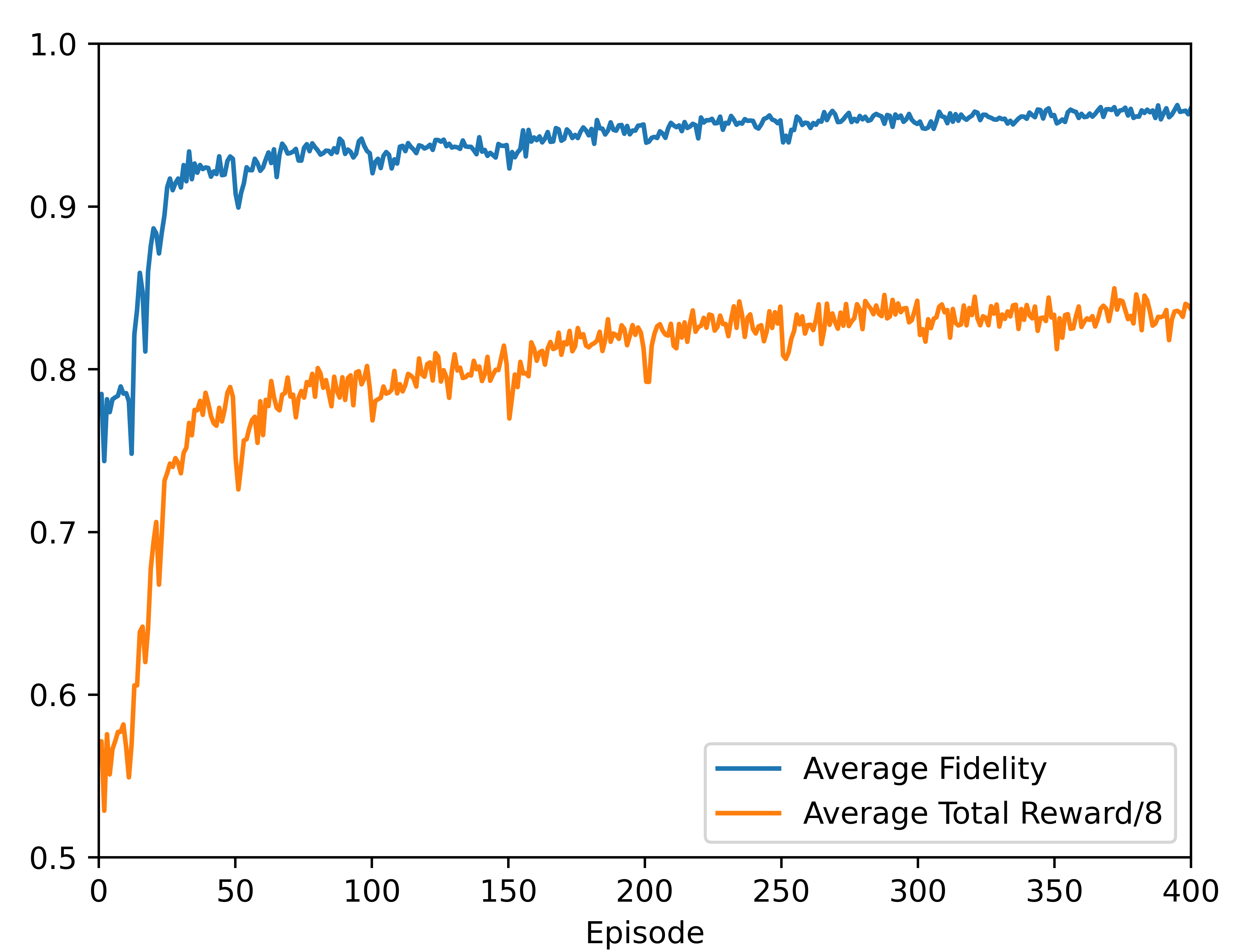}}
		\caption{The average fidelity and total reward over the validation set as functions of the number of episodes in the training process for the two-qubit AQSP.}
		\label{fig:3}
	\end{figure}

	As shown in Fig.~\ref{fig:3}, the average fidelity and total reward value increase rapidly after the first $40$ training episodes, then begin to increase slowly, and the Main Network converges after $280$ training episodes. It is worth mentioning that in order to reduce the pressure on the server memory during the training process, we use a step-by-step training method. Specifically, after every 50 episodes of training, we temporarily save the Main Network and Target Network parameters, and then release the data in the memory. At the next training we will reload the saved network parameters and train further. The Experience Memory $D$ that stores the experience unit is also emptied as the data in the memory is released, so the average fidelity and average total reward value will fluctuate at the beginning of each training session. This fluctuation decreases with the increase of training episodes, and disappears when the Main Network converges. 
	
\begin{figure}
		\centerline{\includegraphics[width=1.0\columnwidth]{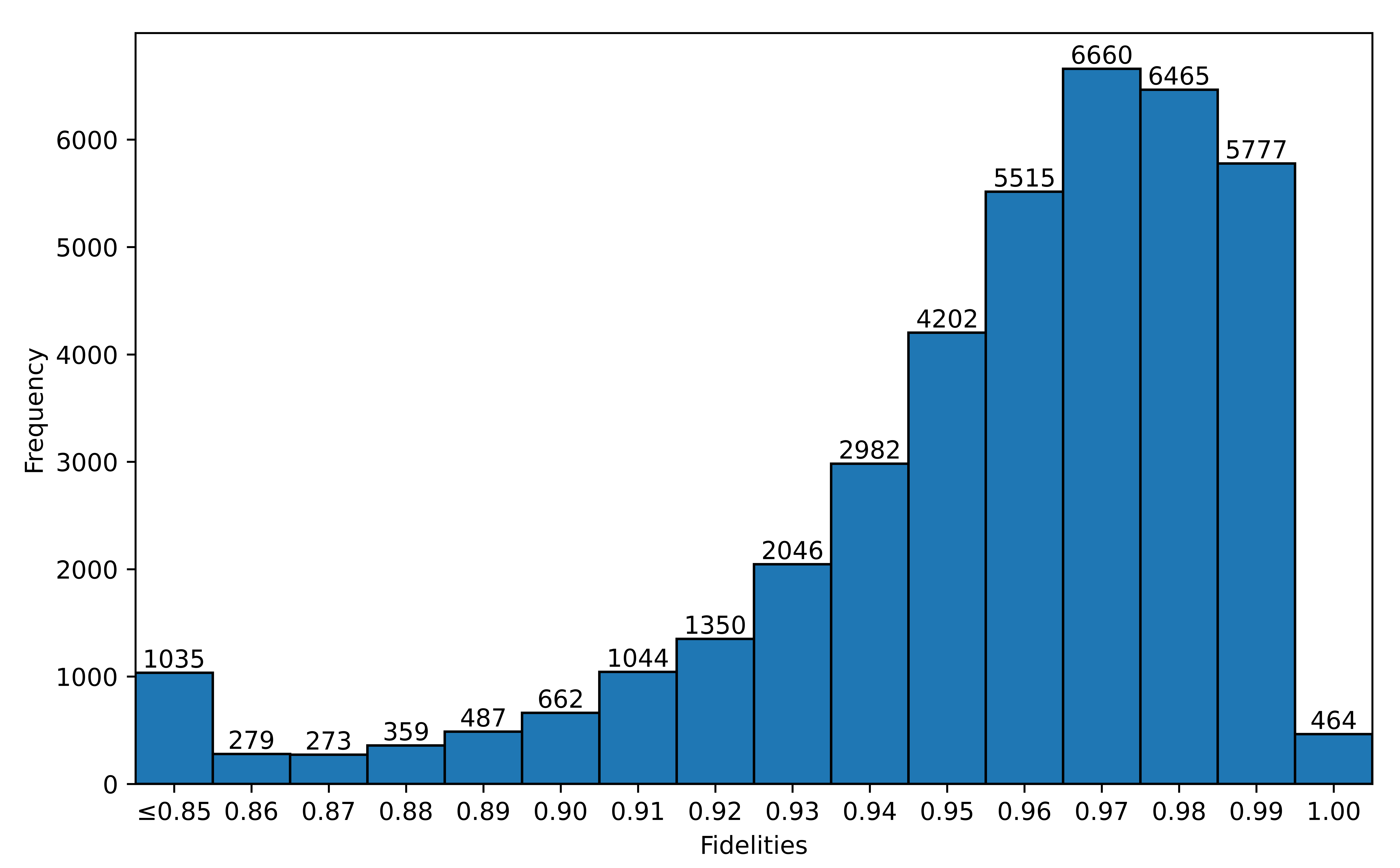}}
		\caption{The fidelity distribution of 39600 samples of the test set in the two-qubit AQSP, the average fidelity is $\bar{F}=0.9556$.}
		\label{fig:4}
	\end{figure}

	After $400$ training episodes, we assessed the Main Network's performance using data from the test set, recording an average fidelity of $\bar{F}=0.9556$. Fig.~\ref{fig:4} illustrates the fidelity distribution for the two-qubit AQSP in the test set, under the control trajectory designed by the AQSP algorithm. The results indicate that the fidelity for the majority of tasks surpasses $0.95$, signifying that the overall performance of the Main Network is commendable.

	\subsection{AQSP in noisy environments}

	In the discussed AQSP tasks, the influence of the external environment was not considered. However, in practice, quantum systems are inevitablely disturbed by its surroundings, which can significantly hinder the precise control of the quantum system. We now proceed to evaluate the performance of the control trajectories designed by the AQSP algorithm in the presence of noise. For the semiconductor DQDs model, it mainly has two kinds of noises: charge noise and nuclear noise. Charge noise primarily originates from flaws in the applied voltage field, while nuclear noise is mainly attributed to uncontrollable hyperfine spin coupling within the material \cite{dots5coherent,roloff2010electric,barnes2012nonperturbative,nguyen2011impurity}.

	For a single-qubit system, these two types of noise can be modeled by introducing minor variations, $\delta\sigma_z$ and $\delta\sigma_x$, into the Hamiltonian in Eq.~(\ref{equ:1}). In the case of a two-qubit system, $\delta_i\sigma_z$ and $\delta_i\sigma_x$ are incorporated into the Hamiltonian to account for the noise effects, where $i$ belongs to the set $\{1,2\}$, representing each qubit, and $\delta$ signifies the noise amplitude. These noise factors are superimposed on the system's evolution after the Main Network has formulated a control trajectory. Specifically, we introduce random intensity noise to various actions within a control trajectory, which is a plausible assumption given the often unpredictable nature of environmental impacts.

\begin{figure}
	\centerline{\includegraphics[width=1.0\columnwidth]{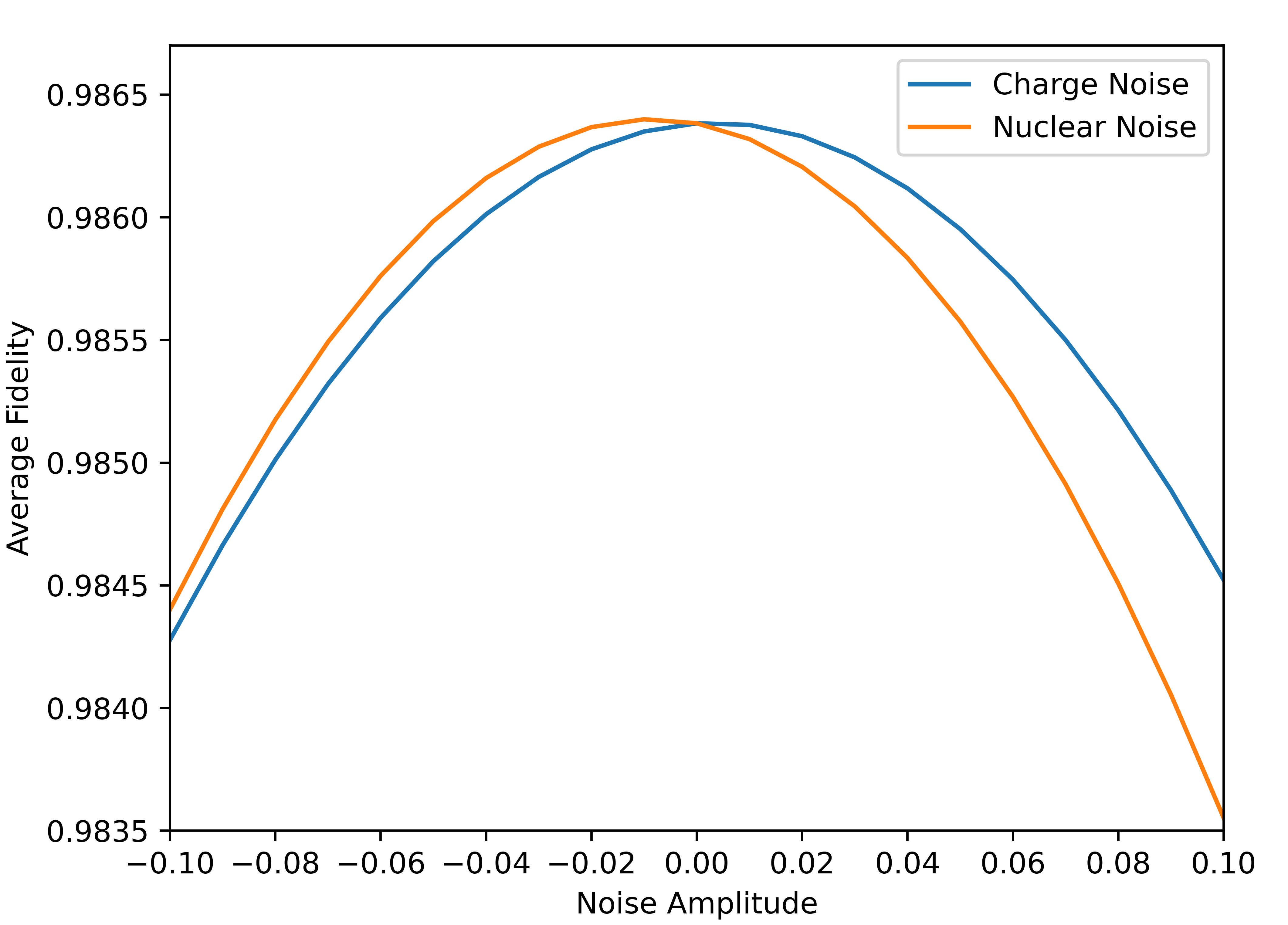}}
	\caption{The average fidelity over the test set as functions of amplitudes of charge and nuclear noises for single-qubit AQSP.}
	\label{fig:5}
	\end{figure}

\begin{figure}
	\centerline{\includegraphics[width=1.0\columnwidth]{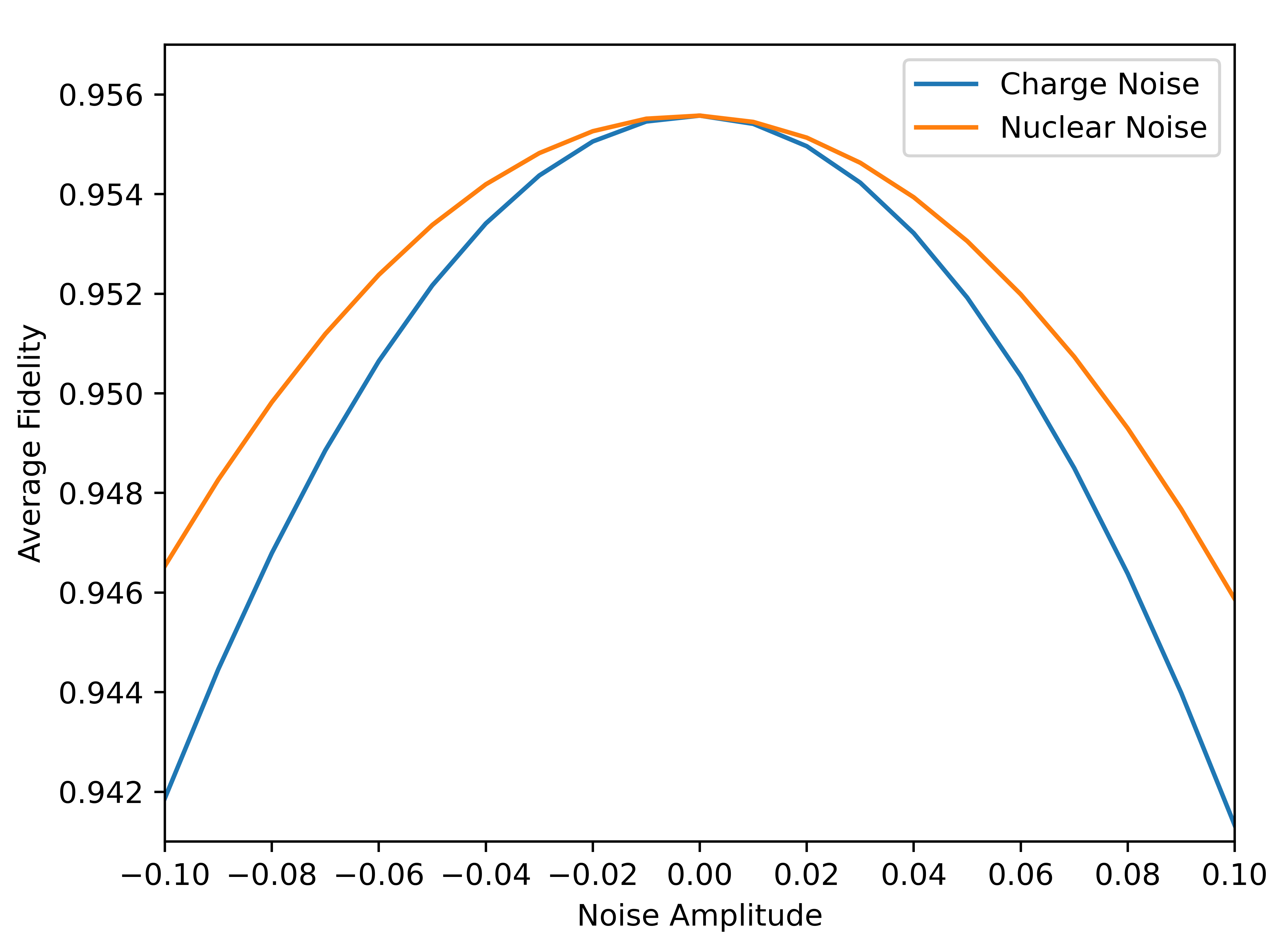}}
	\caption{The average fidelity over the test set as functions of amplitudes of charge and nuclear noises for two-qubit AQSP.}
	\label{fig:6}
	\end{figure}

	Fig.~\ref{fig:5} and Fig.~\ref{fig:6} depict the average fidelity of the preparation tasks for single-qubit and two-qubit arbitrary quantum states, respectively, for the control trajectories generated by the AQSP algorithm under varying noise amplitudes within the test set. It is observable that the average fidelity of the test set remains relatively stable, suggesting that the control trajectory designed by our algorithm exhibits commendable robustness within a certain range of noise amplitude intensity.

	\section{CONCLUSION}

	In this paper, we have effectively designed a control trajectory for the AQSP. This was accomplished by integrating the initial and target state information into a unified state within the architecture of the DQN algorithm. To overcome the challenge posed by the intricate nature of quantum state elements, which are typically not conducive to machine learning applications, we have implemented the POVM method. This approach allows for the successful incorporation of these complex elements into our machine learning framework.
	
	We have assessed the efficacy of the designed control trajectories by testing them on the AQSP for both single- and two-qubit scenarios within a semiconductor DQDs model. The average fidelity achieved for the test set in the preparation of single-qubit and two-qubit arbitrary quantum states are $\bar{F}=0.9864$ and $\bar{F}=0.9556$, respectively. Additionally, our findings indicate that the control trajectories have substantial robustness against both charge noise and nuclear noise, provided the noise levels remain within a specific threshold.  Although our current focus is on state preparation, the proposed scheme possesses the versatility to be extended and applied to a diverse array of multi-objective quantum control challenges.

	\section{ACKNOWLEDGMENTS}
	This paper is supported by the Natural Science Foundation of Shandong Province (Grant No. ZR2021LLZ004) and Fundamental Research Funds for the Central Universities (Grant No. 202364008).
	\bibliographystyle{unsrt}
	\bibliography{References_library.bib}
	\clearpage

	\appendix
	
	\section{Deep reinforcement learning and deep Q network}		
	\setcounter{equation}{0}
	\renewcommand\theequation{A.\arabic{equation}}

	In this section, we detail the DRL and DQN algorithm, which constitute the core of our AQSP framework. DRL is an amalgamation of deep learning and reinforcement learning techniques. Deep learning employs multi-layered neural networks to discern features and patterns from intricate tasks \cite{lecun2015deep}. In contrast, reinforcement learning is a paradigm where a learning Agent progressively masters the art of decision-making to achieve a predefined objective, through continuous interaction with its Environment \cite{sutton2018reinforcement}. Within the purview of DRL, deep neural networks are harnessed to approximate value functions or policy functions, thereby equipping intelligent systems with the acumen to make optimal decisions \cite{shalev2014understanding}.
	
	In the realm of reinforcement learning, an Agent symbolizes an intelligent system that is endowed with the capability to make decisions. The Agent's action selection process is predicated on the Markov decision process framework, wherein an action is chosen solely based on the current state, discounting any past state influences \cite{sutton2018reinforcement}. As the Agent and the Environment engage in a dynamic interaction at a given time $t$, the Agent selects the most advantageous action $a_i$ from a set of possible actions $a=\{a_1,a_2,\cdots,a_n\}$, in response to the Environment's current state $s$, and subsequently executes it. The Environment then transitions to a subsequent state $s^{\prime}$ and bestows an immediate reward $r$ upon the Agent. The Agent employs a policy function $\pi(s) $ to ascertain the most suitable action to undertake, effectively determining $a_i=\pi(s)$.

	A comprehensive decision task yields a cumulative reward $R$, which can be mathematically expressed as \cite{sutton2018reinforcement}:
	\begin{equation}
		R=r_1+\gamma r_2+\gamma^2r_3\cdots+\gamma^{N-1}r_N=\sum_{t=1}^N\gamma^{t-1}r_t,
		\label{equ:6}
	\end{equation}
	where $\gamma$ denotes the discount rate, ranging within the interval $[0,1]$, and $N$ is the total number of actions executed throughout the decision task. The Agent's goal is to maximize $R$, as a higher cumulative reward $R$ signifies superior performance. Owing to the discounted nature of cumulative rewards, the Agent is inherently motivated to secure larger rewards promptly, thereby ensuring a substantial cumulative $R$. To determine the optimal action to take in a given state, we rely on the action-value function, commonly referred to as the $Q$-value \cite{watkins1992q}:
	\begin{equation}
		\begin{split}
			Q^\pi(s,a_i) &=E[r_t+\gamma r_{t+1}+\cdots|s,a_i] \\
			&=E\Big[r_t+\gamma\:Q^\pi\left(s^{\prime},a^{\prime}\right)|s,a_i\Big].
		\end{split}
		\label{equ:7}
	\end{equation}

	\begin{figure}
		\centerline{\includegraphics[width=1.0\columnwidth]{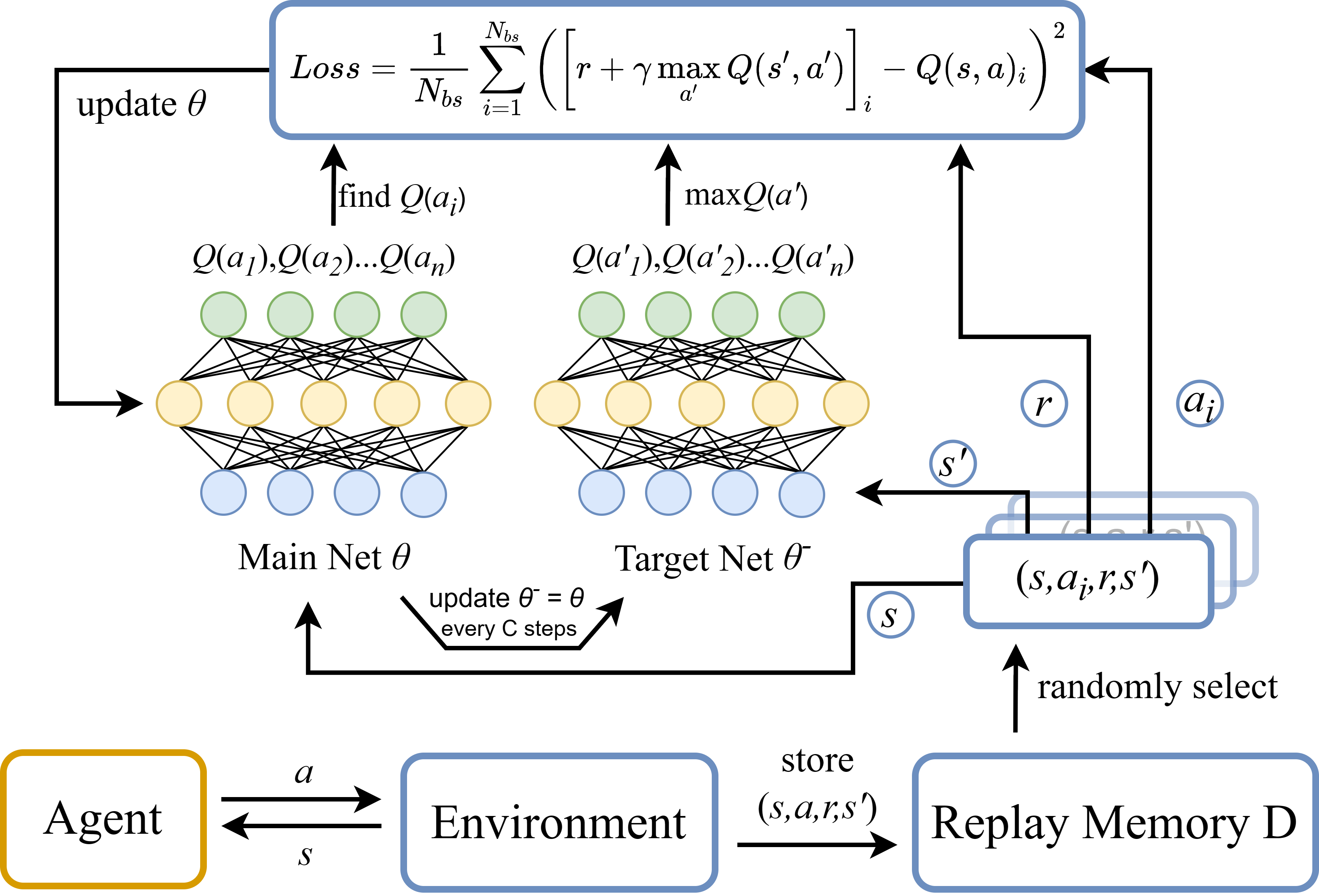}}
		\caption{Schematic for the AQSP algorithm.}
		\label{fig:7}
	\end{figure}

	The $Q$-value embodies the anticipated cumulative reward $R$ that the Agent will garner by executing action $a_i$ in a specific state $s$, in accordance with policy $\pi$. This value can be iteratively computed based on the $Q$-values associated with the ensuing state. In $Q$-learning \cite{watkins1992q}, a $Q$-Table is utilized to record these $Q$-values. Armed with an accurate $Q$-Table, the optimal action to undertake in a given state $s$ becomes readily apparent. The learning process, in essence, revolves around the continual updating of the $Q$-Table, with the $Q$-learning update formula articulated as follows:
	\begin{equation}
		\begin{split}
			Q(s,a_i)\leftarrow Q(s,a_i)+\alpha[r_t+\gamma\max_{a'}Q(s',a')-Q(s,a_i)],
		\end{split}
		\label{equ:8}
	\end{equation}
	where $\alpha$ is the learning rate. When updating the $Q$-value, we consider not only the immediate reward but also the prospective future rewards. The current $Q$-value update for $Q(s,a_i)$ necessitates identifying the maximum $Q$-value $Q(s',a')$ for the subsequent state $s'$, which requires evaluating multiple actions to ascertain the largest $Q$-value. Confronted with this trade-off between exploitation and exploration, we employ the $\epsilon$ -greedy algorithm to select actions \cite{he2021deep}. Specifically, we allocate a probability $\epsilon$ to choose the currently most advantageous action, and a probability of $1-\epsilon$ to explore additional actions. As training advances, $\epsilon$ gradually increases from $0$ to a value just below $1$. This approach enables the $Q$-Table to expand swiftly during the initial phase of training and facilitates efficient $Q$-value updates during the intermediate and final stages of training.
	
	Calculation of $Q$-values for tasks that involve multiple steps and actions can be time-consuming, as the outcome is contingent upon the sequence of actions chosen. To address this challenge, we utilize a multi-layer artificial neural network as an alternative to a $Q$-table. A trained neural network is capable of estimating $Q$-values for various actions within a given state. The Deep $Q$-Network (DQN) algorithm \cite{mnih2013playing, mnih2015human} comprises two neural networks with identical architectures. The Main Net $\theta$ and the Target Net $\theta^-$ are employed to predict the terms $Q(s,a_i)$ and $\max_{a^{\prime}}Q(s^{\prime},a^{\prime})$ from Eq.~(\ref{equ:8}), respectively.
	
	We implement an experience replay strategy \cite{mnih2013playing} to train the Main Net. Throughout the training process, the Agent accumulates experience units $(s, a, r, s')$ at each step, storing them in an Experience Memory $D$ with a memory capacity $M$. The Agent then randomly selects a batch of $N_{bs}$ experience units from the Experience Memory $D$ to train the Main Net. The loss function is calculated as follows:
	\begin{equation}
		Loss=\frac{1}{N_{bs}}\sum_{i=1}^{N_{bs}}\left(\left[r+\gamma\max_{a^{\prime}}Q\left(s^{\prime},a^{\prime}\right)\right]_i-Q(s,a)_i\right)^2
		\label{equ:9}
	\end{equation}	
	where we use Eq.~(\ref{equ:9}) to calculate the Loss and refine the parameters of the Main Net $\theta$ using the mini-batch gradient descent (MBGD) algorithm  \cite{mnih2013playing, mnih2015human}. The Target Net $\theta^-$ remains inactive during the training process, only updating its parameters by directly copying from the Main Net $\theta$ after every $C$ steps. The schematic diagram illustrating the AQSP algorithm is presented in Fig.~\ref{fig:7}.

\end{document}